\documentclass[11pt,twoside]{article}

\usepackage{asp2006}
\usepackage{epsf}
\usepackage{lscape}

\begin{document}

\title{Stellar processes near AGN}   
\author{Sergei Nayakshin}   
\affil{Dept. of Physics \& Astronomy, University of Leicester, LE1 7RH, UK}    

\begin{abstract} 
Precise mechanisms by which Active Galactic Nuclei (AGN) receive their gaseous
fuel is still a mystery. Here I draw attention to the extra ordinary star
formation event that took place in the central $\sim 0.5$ parsec of our
Galaxy. The most reliable explanation of the event seems to be that two
somewhat massive nearly co-eval gaseous disks failed to accrete on Sgr~A$^*$,
the super-massive black hole (SMBH) in our Galaxy, and instead cooled down and
gravitationally collapsed, forming the stars observed now. This emphasises
that star formation must be an important part of AGN feeding puzzle. I also
discuss a model in which stellar winds create the observed obscuration of AGN.
These winds are cold, clumpy and dusty, as required by the observations, but
they are Compton-thin unless wind outflow rate is highly super-Eddington. This
argument is in fact a general one, independent of the wind driving
mechanism. I thus suggest that winds may be important for optically thin
absorbers, and that a better model for optically thick AGN obscuration is a
warped accretion/star forming disk.
\end{abstract}



\section{Introduction}

In X-ray binary systems, we know quite well the origin and properties of gas
arriving at the outer edge of an accretion disk. In contrast to this, we know
nothing for a fact about how gas trickles down into the inner parsecs of
AGN. This is because observations of these regions are tremendously difficult:
one needs to resolve very small angular scales faced at the same time with the
contaminating emission from few tens of gravitational radii of SMBH. Despite
all this, there is now a significant progress in such observations
\citep{Davies06}.

The Galactic Center (GC) is our most suitable laboratory for these studies due
to its proximity and the fact that Sgr~A$^*$ is in fact a very dim source,
producing bolometrically only $\sim$ hundreds of Solar luminosities. As a
result, young massive ``He-I'' stars dominate the power output of the central
parsec of our Galaxy. Many of these stars are now resolved to be located in a
very well defined and rather thin stellar disc that rotates clockwise as seen
on the sky \cite{Levin03,Genzel03,Paumard06}. The rest of the stars can be
arguably classified as a second more diffuse disc or a feature that rotates
counter clock-wise \cite{Genzel03,Paumard06,LuEtal06}.

Below I shall argue that the best interpretation of data on these young
massive stars is star formation inside an accretion disk(s) that existed in
the inner parsec just before the stars were born. I shall discuss some
observational and theoretical issues pertaining to formation and mass spectrum
of these stars. I will also present results of numerical simulations of star
formation in a massive gaseous disc around Sgr~A$^*$.

Using Sgr~A$^*$ observations as a launch pad for a discussion on the state of
gas arriving in the inner parsecs of AGN, I will then argue that in general
AGN disks are star-forming as well. I will then present numerical simulations
of stellar winds escaping from a stellar disc left by a star formation
event. Such winds might provide some obscuration of AGN, as required by
observations. The results of the simulations however show that such outflows
are unable to provide Compton-thick obscuration unless winds carry
significantly super-Eddington outflow rates. A simple analytical estimate
shows that this result is general for any wind-driving mechanism. I thus argue
that, while certainly present and important for the working and appearance of
AGN, {\em outflows are not responsible for Compton-thick obscuration}.
Instead, I propose that the role of the Compton-thick absorber is played by a
geometrically thin but strongly warped accretion/star-formation disk. As an
example, gravitational torques due to a non-spherical stellar potential are
demonstrated to be effective in producing strongly warped gaseous discs.

\section{Observational constraints on the young stars in Sgr~A$^*$ neighbourhood}

\subsection{The in-situ origin of the stars}

There is nearly a hundred early type stars in the central parsec of our GC
\citep{Paumard06}. Finding these young stars there, as close as $\sim 0.03$ pc
away from a SMBH, was a real surprise.  ``Standard'' models of star formation
are not easily applicable here due to a huge tidal field of the central object
at $R= 0.1$ pc distances from Sgr~A$^*$.  The required gas density is $n_H >
10^{11} \hbox{cm}^{-3} (R/0.1\hbox{pc})^{-3}$, i.e., exceedingly large
compared with gas densities in galactic molecular clouds.

It thus has been suggested that the stars may have been formed at a distance
of tens of parsecs \citep{Gerhard01}, avoiding the need for the excessive gas
density prior to star formation, in a massive star cluster. The cluster's
orbit would then decay through dynamical friction with the background stars.
However, \cite{Paumard06} find a well defined outer edge of about $\simeq
0.5$~parsec to the observed distribution of young massive stars in the
GC. This is difficult to understand if stars migrated into the inner parsec
from outside, as the disrupted cluster should leave a trail of stars behind
it.

If the cluster were very strongly mass-segregated, i.e., that massive stars
were present only in its core, then the trail could be composed of low mass
stars only. These stars could escape detection in the near infra-red. This
model seems to be challenged by X-ray data. Recent calibration of X-ray
properties of T Tauri stars by \cite{PreibischEtal05} revealed that they are
very bright in X-rays, i.e. some three orders of magnitude brighter than they
are on the main sequence. They would be observable even at high extinction
regions such as the GC. These young stars also display giant X-ray flares,
with some becoming as bright as $\sim 10^{33}$~erg~s$^{-1}$. \cite{NS05} used
these new X-ray results for young stars, and observations of the GC reported
in \cite{Muno04}, to show that the inner $\sim 10$ parsec could not be hiding
more than $10^4$ or so young low mass stars. This is insufficient: some $ 10^5
- 10^6 $ Solar masses of young stellar mass is needed to make the cluster
heavy enough for it to sink in during the short lifetime of the young massive
stars.

\subsection{Top-heavy mass spectrum of the in-situ star formation
  event}\label{sec:topheavy} 

Because Sgr~A$^*$ is so dim in X-rays, it turned out possible to push the
X-ray constraints further. Baganoff et al. (2003) described the properties of
the unresolved X-ray emission near Sgr~A$^*$ in detail. Nayakshin \& Sunyaev
(2005) used these results to deduce that the area most densely populated by
massive young stars, i.e., the inner $R \le 0.2$ parsec could contain no more
than $\sim 10^3 $ Solar masses of low-mass YSO ($M < 3$ Solar mass).  This is
interesting since a factor of 10 or so more would be expected if the IMF were
the ``normal'' galactic one, such as \cite{Miller79}.

\subsection{Total stellar mass created}\label{sec:orbital} 

An important question to ask is the total mass budget of the star formation
event. The best age estimate for the young stars in the GC is $\sim 6\pm 1$
million years and their combined mass is a few thousand Solar mass
\citep{Paumard06}. Due to the age of this stellar population, stars more
massive than $\sim 50-60$ Solar masses seem to be absent. Due to the
top-heavy nature of the mass spectrum of the young stars discussed above, it
is then possible that the initial star formation event created a star cluster
far more massive than the one we observe now.

\cite{Nayakshin05,NC05} proposed a way to constrain the total stellar mass
using observational constraints on stellar orbits. Suppose that the two
stellar systems were created infinitely thin and flat, as they would be in the
simplest self-gravitating disc scenario (see \S 3 below). With time, an
isolated stellar disc will thicken due to internal $N$-body heating, and two
stellar discs will warp each other due to their non-spherical gravitational
potentials. Both of these effects are stronger the more massive the stellar
discs are. Now, if the initial systems were not thin and flat, then the
thickening and warping will be even quicker
\citep[e.g.,][]{NDCG06}. Therefore, demanding that the model orbital
configuration fits two stellar planes no worse than the real stars do at the
present moment, we can arrive at a constraint on the total mass of these
discs.  This yields the limit of around $10^4$ solar masses for the total
masses of each of the stellar discs.

\section{Theoretical models}\label{sec:theory}

It has been known for a long time that massive accretion discs can be
gravitationally unstable and form stars if they are able to cool efficiently
\citep[e.g.,][]{Paczynski78,Gammie01,Rice05}. In particular, the disc has to
be both massive enough and be able to cool fast enough: Toomre parameter
$Q\le1$, and $t_{\rm cool} < 3 \Omega^{-1}$, where $\Omega = \sqrt{GM_{\rm
BH}/R^3}$.  Vertically integrated marginally star-forming disc models have
been considered recently with application to Sgr~A$^*$ in mind
\citep{Levin06,Nayakshin06}. Pleasingly, these models have a number of
features consistent with the observations. The inner edge of the
self-gravitating region, i.e. the closest to Sgr~A$^*$ the stars can be born
in this scenario, is about $0.03$ pc, consistent with the distribution of the
disk-stars \citep{Paumard06}. Further, the models predict that all the gaseous
disk mass would go into building up stars. The stellar mass is then estimated
to be at $\sim 10^4$ Solar masses, which is not inconsistent with the
observations.

The issue of the top-heavy mass function is more interesting. The estimates
for the fragmentation mass, $M_{\rm frag}$, i.e. the mass of the first bound
fragments, are certainly below 0.1 Solar mass.  Hence, if disk were to {\em
promptly} collapse into clumps of mass of this order, one would expect
low-mass stars to dominate the mass spectrum of collapsed objects.  However,
in reality, collapse of gas clumps may not be dynamical (as assumed by the
simple model), but gradual, regulated by cooling and clump rotation.  Further,
the clumps will have sizes comparable to the dimensions of the ``first
cores'', i.e. $R_{\rm clump} \sim 5$~AU. The clumps are actually likely to
merge and grow by agglomeration \cite{Levin06}. Therefore one expects that
$M_{\rm frag}$ is a strong under-estimate of the actual stellar mass.
Further, Nayakshin (2006) noted that low-mass proto-stars born with $M\sim
M_{\rm frag}$ may grow very massive by gas accretion if disc fragmentation is
slow. In particular, stellar accretion luminosity, produced by newly born
proto-stars, turns out to be sufficient to heat the disc up, increasing the
Toomre $Q$-parameter of the disc above unity and hence making the disc stable
to further fragmentation. Numerical simulations (see below) do bear this
prediction out.

\section{Numerical models of star-forming discs}

We use the SPH/$N$-body code {\small GADGET-2} \cite{Springel05} to simulate
the dynamics of stars and gas in the (Newtonian) gravitational field of
Sgr~A$^*$. Details of our method will be reported in Nayakshin, Cuadra \&
Springel 2007.  Radiative cooling of the disc is treated with a simple locally
constant cooling time prescription, $t_{\rm cool} = \beta/\Omega$, where
$\beta$ is a constant of order unity.

\begin{figure}[h]
\plottwo{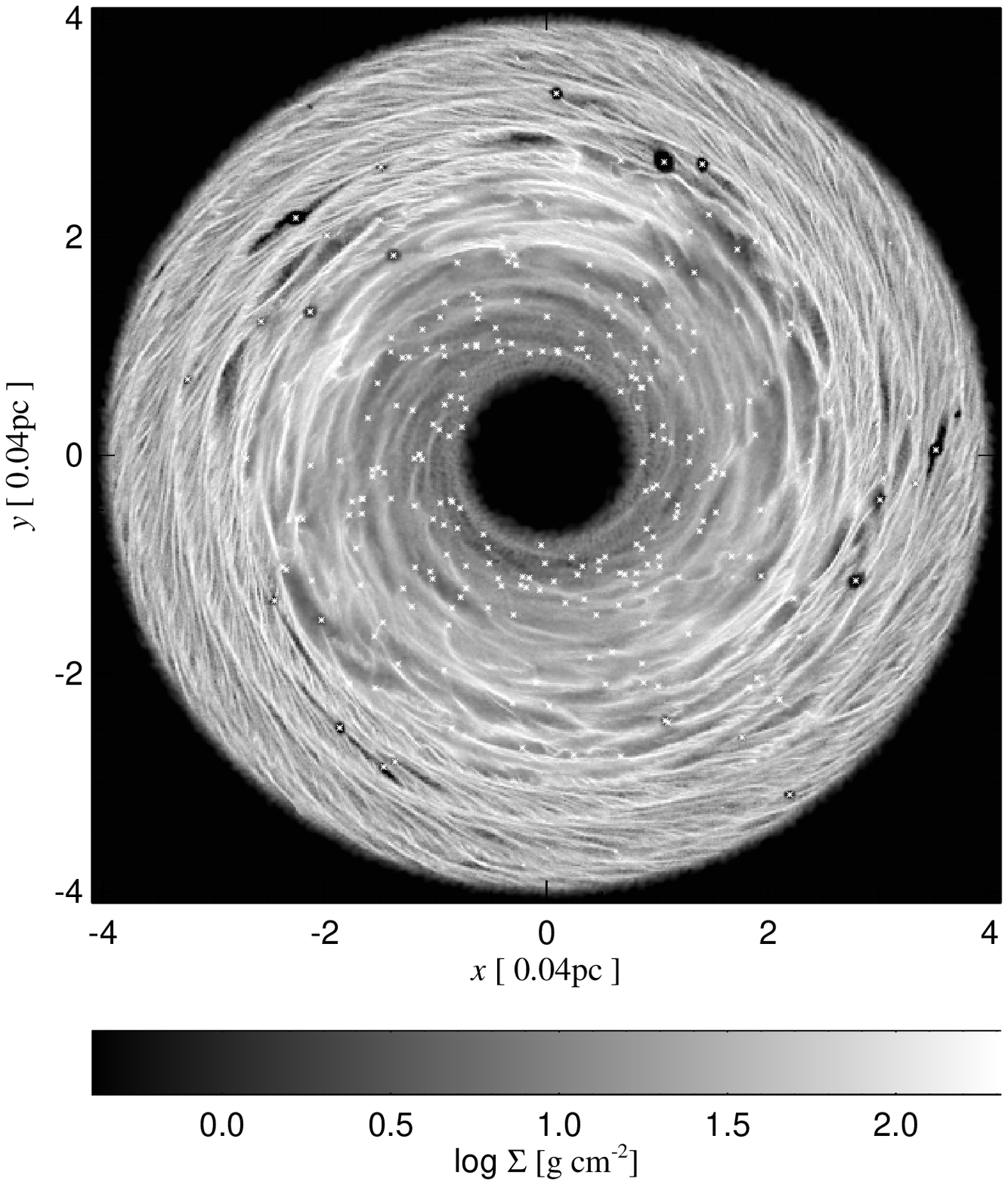}{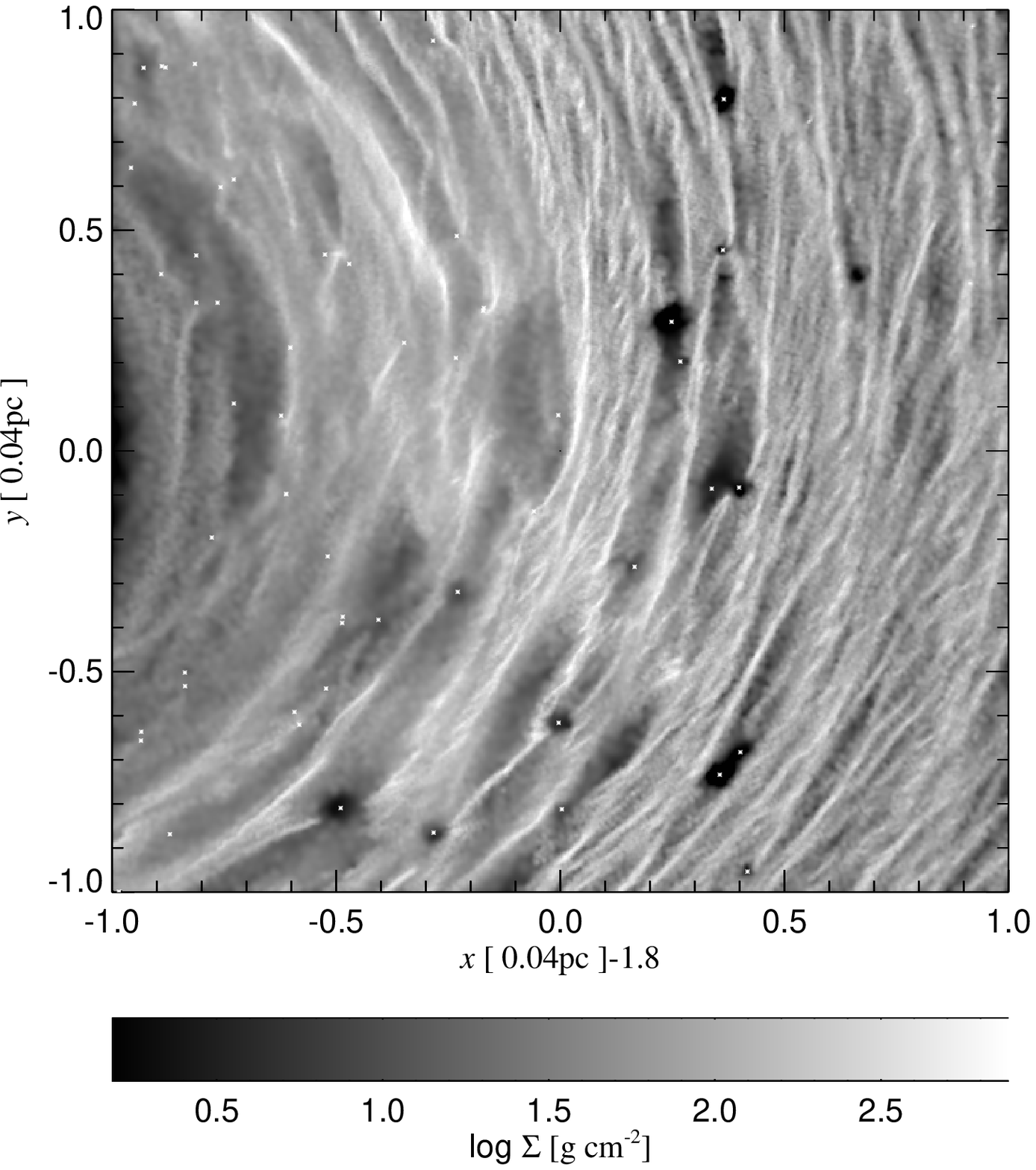}
\caption{Left panel: Snapshot of a star-forming disc of initial mass $2\times
  10^4$ Solar masses at time $t=10,000$ years. Stars are shown as
  asterisks. Right panel: Same shown for a small patch of the disk at time
  $t=7,000$ years.}
\label{fig:late}
\end{figure}

\paragraph{\bf Circular gas discs without feedback.}

Left panel of Figure \ref{fig:late} shows a snapshot of a run with $\beta=3$
well into the non-linear stage, at time $t\approx 10^4$ years, when more than
half of the gas was already turned into stars.  The initial condition is a gas
disc of mass $2\times 10^4 $ Solar mass in Keplerian circular rotation around
Sgr~A$^*$, extending from 1'' to 4'' (1''$\approx 0.04$~pc). The right panel
shows the same simulation in a slightly earlier time but now zoomed into a
smaller patch of the disk to emphasise the disk structure.

Star formation is fastest in the innermost region, where most of the gas is
already depleted. At the end of the simulation essentially all the gas is
turned into stars. Unlike ``normal'' star formation, feedback from massive
stars will not be very effective in blowing the gas away via radiation
pressure or winds as the stars are within the deep potential well due to
Sgr~A$^*$. The stars thus steal the majority of SMBH's dinner.

\begin{figure}[ht]
\plottwo{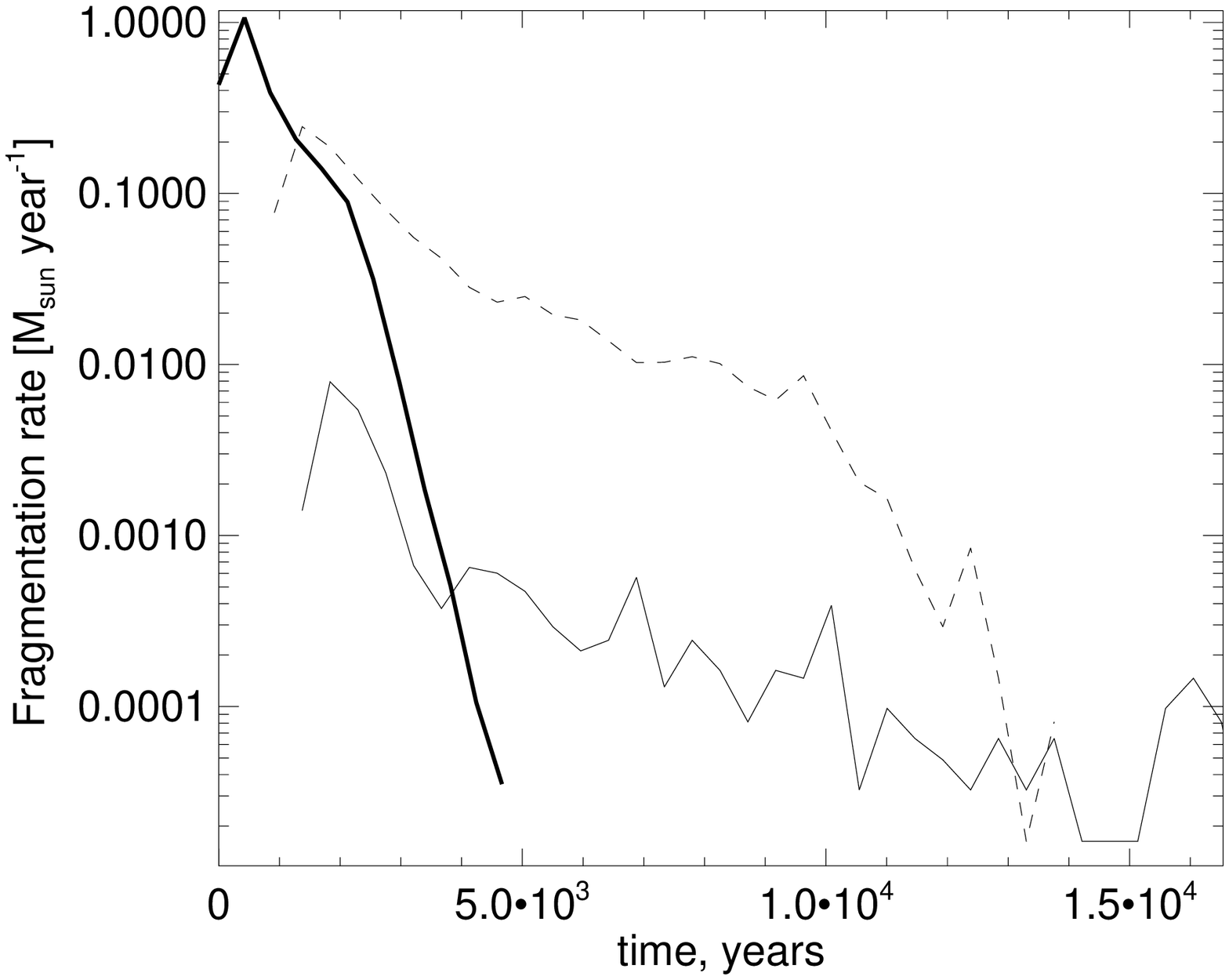}{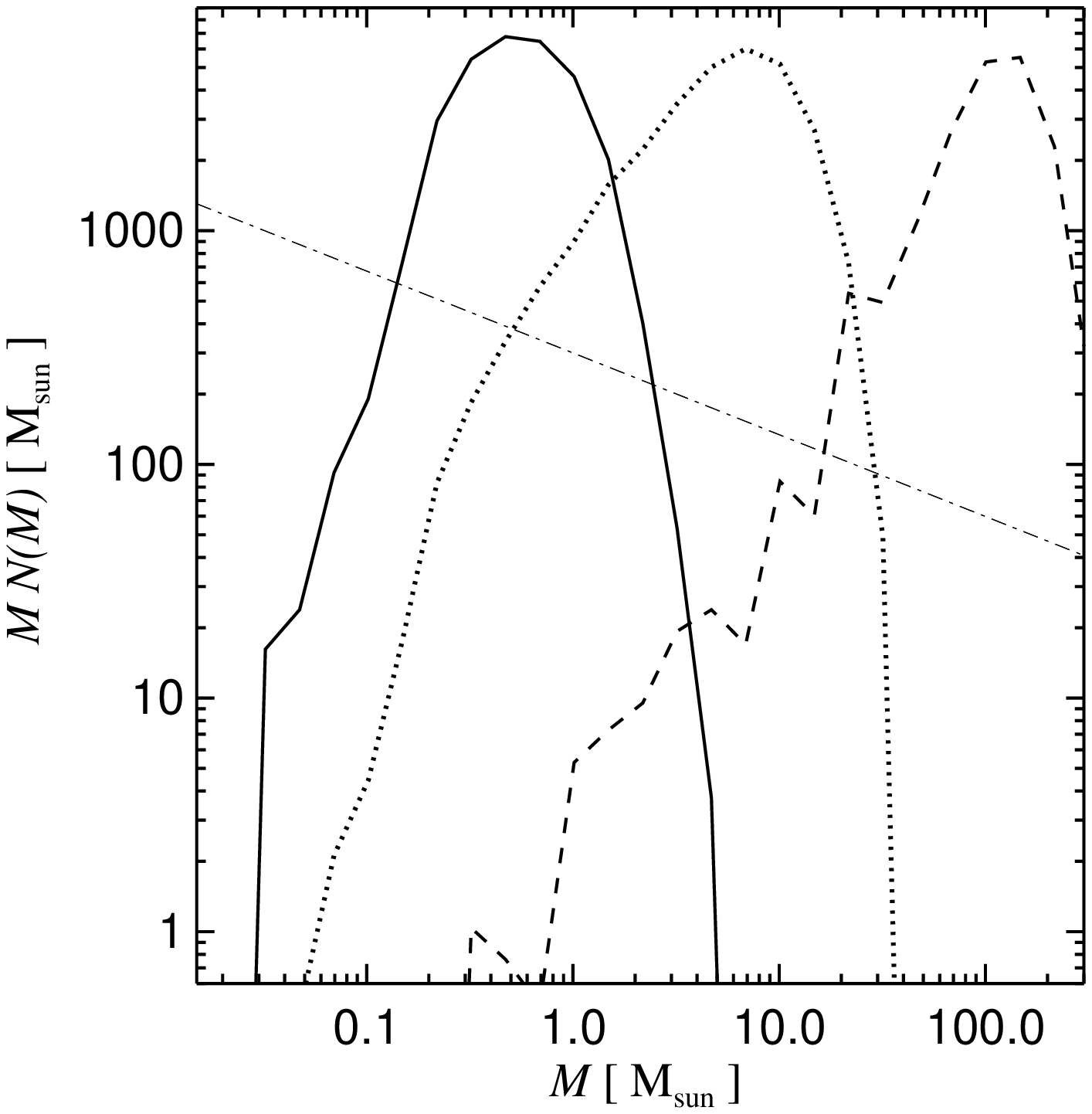}
\caption{Left panel: Disc fragmentation rate in the tests with $\beta=0.3$
  (thick solid), $\beta=2$ (dashed), and $\beta=3$ (thin solid). The shorter
  the cooling time, the faster disc fragmentation proceeds. Right panel: Mass
  function of stars (``IMF'') formed in the simulations with $\beta=0.3$
  (solid), $\beta=2$ (dotted), and $\beta=3$ (dashed). The dot-dashed
  power-law is the Salpeter IMF.}
\label{fig:imf}
\end{figure}

Interestingly, gravitational heating generated by stars (scattering off each
other and interacting with gas) is sufficient to heat the disc up above its
pre-star-formation value, slowing down and even shutting off fragmentation at
later times. The effect is more pronounced the longer the cooling time.
Figure \ref{fig:imf}, left panel, shows fragmentation rate in Solar masses per
year as a function of time for simulations that differ only by the cooling
time parameter $\beta$, as labelled in the caption. The longer the cooling
time, the sooner disc fragmentation slows down and eventually stops as the
disc becomes hotter. The total mass budget of the simulation is fixed by the
initial condition. Quite logically then, the longer the cooling time the more
top-heavy the mass function of the stars born in the simulation becomes. The
IMF of the stars from these three simulations is presented in Figure
\ref{fig:imf}, right panel.

\paragraph{\bf The role of feedback on the fate of the discs and the IMF.}

As already mentioned, Nayakshin (2006) suggested that the accretion luminosity
of young stars within a star-forming disk might be strong enough to warm it up
and significantly slow down its fragmentation. We have made several numerical
runs with stellar accretion feedback included. As expected, we found a very
strong reduction in the fragmentation rate, and an increase in the disk
lifetime defined as the time it takes for a half of its initial mass to be
turned into stars. We could not reach a conclusive statement on the influence
of the stellar accretion luminosity feedback onto the stellar mass function as
the simulations could not be run sufficiently long. 

The fact that disk lifetime increases strongly in the simulations is
encouraging in terms of satisfying SMBH growth needs. If these results could
be extrapolated into a much heavier accretion disc case, the time scale for
angular momentum transfer mediated by self-gravity could become shorter than
the disc lifetime. Unfortunately, this appears unlikely. In the simulations we
used the simple $\beta$ cooling time prescription. Quite a general analytical
argument shows that, in order to feed AGN at rates of order Eddington
accretion rates, the required levels of energy feedback from star formation
(and in fact of any extra disc heating at all) are so large that they strongly
contradicts the observed spectra of a typical AGN \citep{Goodman03}.

\paragraph{\bf Star formation in eccentric discs} The second of the observed
stellar discs in the inner parsec of our Galaxy is a not so well defined
diffuse feature with stellar orbits that appear to be rather eccentric
\citep[$e\sim 0.5-0.8$][]{Paumard06}. Since N-body relaxation time scale is
about a Gyr, these stars must have been born on already eccentric orbits. We
performed a series of tests aiming at establishing whether star formation can
take place in eccentric discs. The results are rather similar to the runs with
circular gas discs. 

We think that the best case scenario for the GC stellar populations is one
where the clock-wise disc stars formed out of a more massive circular gaseous
disc, whereas the counter clockwise stars were born out of an eccentric stream
of gas. Orbital precession could then turn an initially flat eccentric disc
into a geometrically thicker one by now.

\section{Star-forming winds and AGN obscuration}

The talk by Moshe Elitzur \citep{Elitzur06} summarises the observational need
for a dusty AGN absorber and reviews important theoretical issues in that
topic. It is proposed that a dusty clumpy outflow might be the best
theoretical explanation for the absorber. Here I discuss related ideas. In
particular, star formation on sub-parsec scales, discussed in previous
sections of this paper, will naturally lead to stellar outflows. Provided the
mass outflow rate is large enough, winds will be dense and thus cold, and
quite likely dust-rich. In this case one might expect these outflows to play
an important role in the obscuration of AGN.

SPH numerical modelling of stellar wind accretion on Sgr A$^*$ was recently
developed by us (Cuadra et al. 2005,2006). Nayakshin \& Cuadra (2007) used
that numerical approach to model stellar wind obscuration, concentrating on
higher mass outflow rates than in Sgr A$^*$ presently, as that might be
expected in a bright AGN environment.

In the simulation presented here, stellar mass loss rates are $2.5 \times
10^{-4} $ Solar mass year$^{-1}$ per star. In total, we have 200 mass
shredding stars, thus amounting to the mass loss rate of 0.1 Solar mass per
year, which is a factor of few super-Eddington for Sgr A$^*$ mass of $M = 3.5
\times 10^6$ Solar mass. The stars are situated in a flat circularly rotating
Keplerian disk of geometrical thickness $H(R) = 0.1 R$, where $R$ is the
radius\footnote{The unit of length used here is 1 arcsecond, which corresponds
to about $1.2\times 10^{17}$~cm or 0.04 pc at the Galactic Centre distance of
8 kpc.}.  The disc inner and outer radii are $R_{\rm in} = 1.5$ and $R_{\rm
out} = 8$, respectively.  

\begin{figure}[ht]
\plottwo{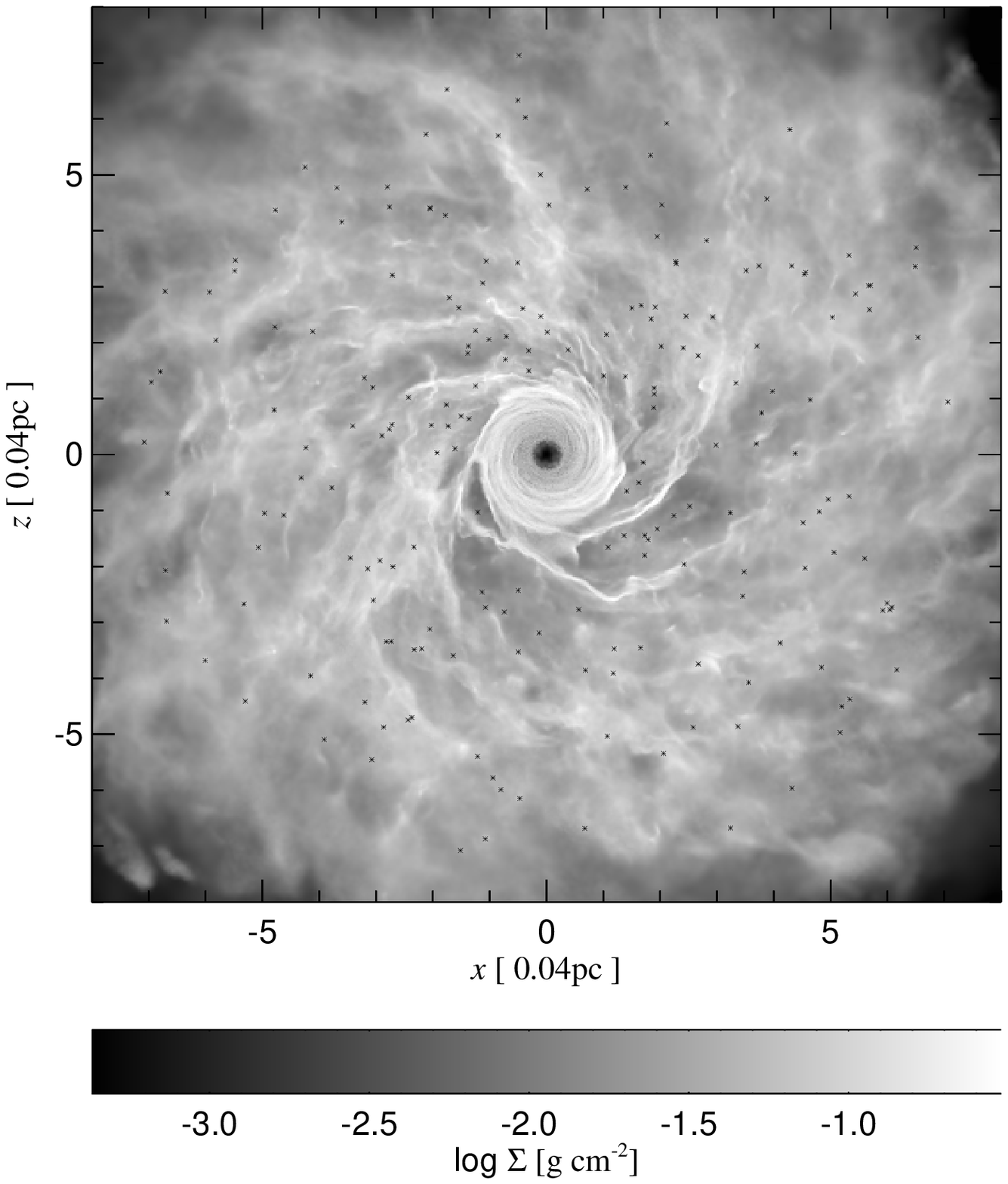}{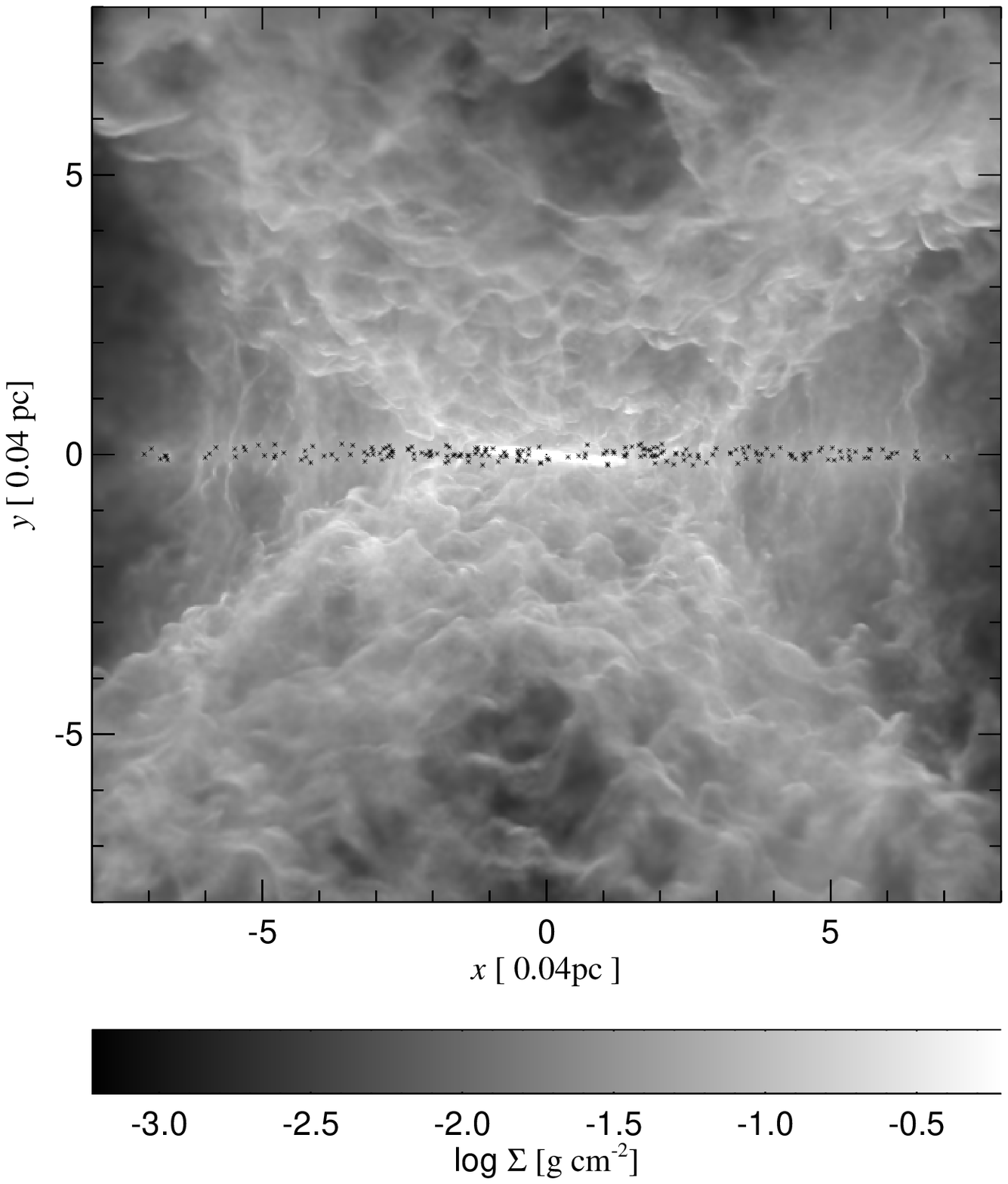}
\caption{Face-on (left panel) and edge-on (right panel) views of the
  simulation domain for the flat stellar disc configuration. Black asterisks
  show the location of the wind-producing stars used in the simulation. From
  Nayakshin \& Cuadra (2007)}
\label{fig:wind1}
\end{figure}

A snapshot of the simulation is shown in Figure \ref{fig:wind1} at time $t
\approx 2200$ years after the beginning of the simulation. While there is no
true steady state for a system of a finite number of moving stars, the
snapshot is fairly typical of the morphology of the stellar wind. The face-on
view (left panel) shows that some of the shocked wind managed to cool down and
formed a small-scale disc.  Smaller escape velocity and less frequent shocks
at the larger radii allow direct escape of both the fast diffuse and the
slower cooler clumpy winds.  Due to the final extent of the stellar disc and a
projection effect, the wind morphology reminds an ``X''-shape
(Fig.~\ref{fig:wind1}, right panel).

\begin{figure}[ht]
\plottwo{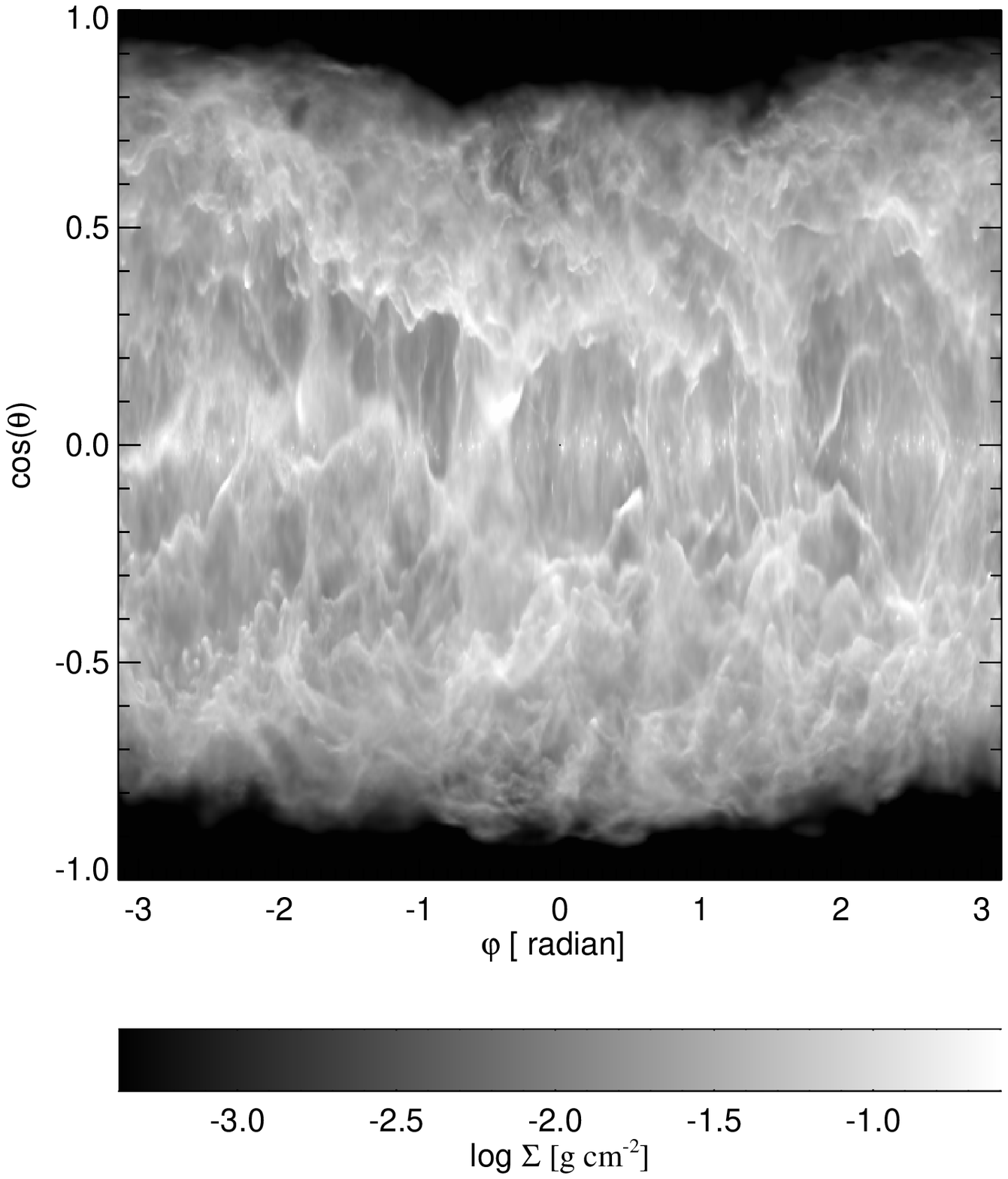}{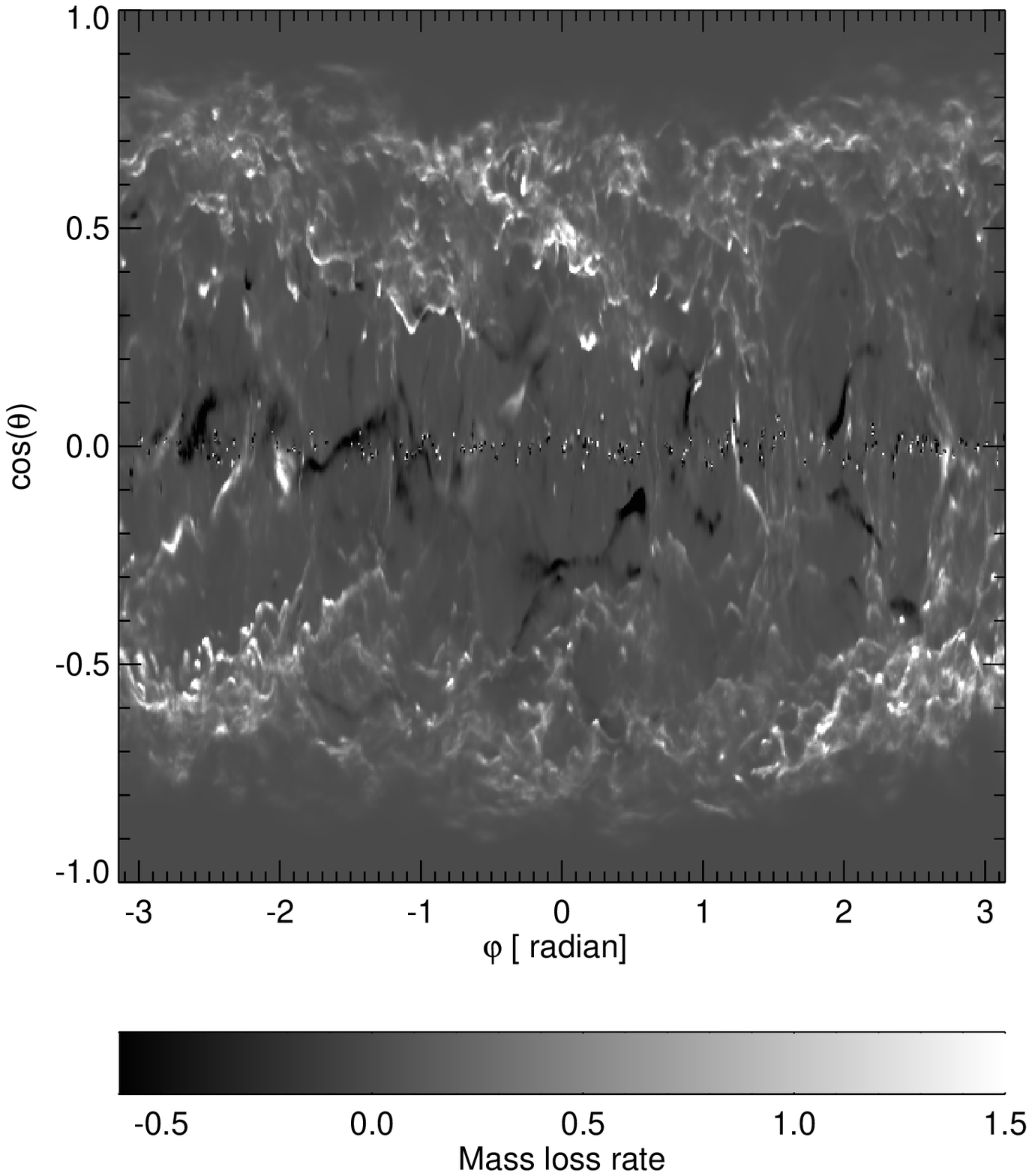}
\caption{Column depth through the wind (left) and isotropic mass outflow rate
  (right; in units of Solar mass year$^{-1}$) for the simulation shown in
  Fig.~\ref{fig:wind1}, as seen from the SMBH.  Note the large variations of
  the obscuring column depth over small angular scales. The gaseous inner disc
  seen in the left panel of Fig.~\ref{fig:wind1} has been excluded.}
\label{fig:wind2}
\end{figure}

The focus of our attention here is on the obscuration properties of these
winds, and not on that of gaseous discs. The left panel of Figure
\ref{fig:wind2} shows the obscuring column depth of the winds as seen from the
SMBH, excluding the gas with radial distances $R < 1.6$ from Sgr~A$^*$, which
removes most of the gaseous disc. Notice the very irregular patchy structure
of the (brighter) optically thicker regions. The contrast between those and
neighbouring less dense patches of sky is frequently a factor of 10 or
more. The dotted pattern at the $\cos \theta= 0$ plane are the dense regions
of stellar winds immediately next to the stars.

The right panel of Figure \ref{fig:wind2} shows the isotropic mass loss rate
along the line of sight, defined as
\begin{equation}
\dot M \equiv \frac{\int d \Sigma \; 4 \pi r^2 \rho v_R}{\int d \Sigma}\;,
\label{defmdot}
\end{equation}
where the integral is taken along the line of sight defined by a given
$\theta$ and $\phi$. 

Now, while the obscuration pattern in the left panel of Figure \ref{fig:wind2}
may certainly be relevant to the observational need for a cold dusty absorber,
there appears to be a serious limitation for this model. The average column
depth is not large enough, i.e. it is Compton-thin. However, observations
require a large fraction of AGN to be Compton-thick
\citep[e.g.,][]{Sazonov04,Guainazzi05}. To satisfy this requirement, the mass
outflow rate, already super-Eddington by a factor of $\sim$ 3, needs to be
pushed another factor of $> 5$ higher. 

This is an entirely general point. Considering clouds with average column
depth $\Sigma_{\rm c}$ outflowing at speeds about the escape velocity, and
numbering $N_{\rm a}$ per line of sight on average, one arrives \citep{NC07}
at the mass outflow rate of
\begin{equation}
\dot M_{\rm wind} \sim 15 \frac{M_{\odot}}{\hbox{year}}\; N_{\rm a}
\Sigma_{\rm c} \left(\frac{r_{\rm t}}{\hbox {1 pc}}\right)^{1/2} M_8^{1/2} \;,
\label{mwreq}
\end{equation}
where $M_8$ is the SMBH mass in units of $10^8 M_{\odot}$, and $r_t$ is the
``torus size''. Now, the Eddington accretion rate is $4\pi G M m_{\rm
p}/\epsilon c \sigma_{\rm T} \approx 2 M_{\odot} \hbox{year}^{-1} M_8$. The
outflow is Compton-thick when $N_{\rm a} \Sigma_{\rm c}\ge 1$. Thus the
required wind mass loss rate (equation \ref{mwreq}) is an order of magnitude
higher, typically, than the Eddington accretion rate for a $M_8 \sim 1$ object
and a parsec-scale torus.

Considering a specific case of the local obscured AGNs studied by
Guainazzi~et. al. (2005), we note that the bolometric luminosities of these
objects in the infrared, X-ray and optical bands are in the range $L \sim
10^{43} - \hbox {few} \times 10^{44}$ erg/sec, which implies SMBH accretion
rates of ``only'' $ \sim 0.01 M_{\odot}$~year$^{-1}$ for the standard
radiative efficiency. Hence if the obscuration of the optically thick objects
in that sample were provided by the winds, we would conclude that the SMBH
accretion process must be very wasteful, with $\sim 100-10,000$ times more
mass flowing out of the inner parsec than accreting on the SMBH. It would also
require a very high mass influx into the inner parsec to sustain such
winds. Given the difficulty of delivering enough fuel to the SMBHs even in the
earlier gas-rich epochs, it is hard to see how such high mass in-fluxes could
be maintained in the local AGN.

\section{Warped accretion disc as an effective absorber}

Given unrealistically high mass outflow rates required to obscure AGN in the
wind outflow model, I suggest that outflows, while certainly being important
for obscuration in optically thin AGN, do not nevertheless provide the bulk of
the obscuration that must be optically thick. This is not to say that winds
are not important for AGN physics!

I argue that a better physically motivated model for the absorber would be a
warped accretion disc. The disc may be significantly warped by instabilities
due to back reaction to mass outflow, or due to the AGN radiation pressure
\citep[e.g.][]{Schandl94,Pringle96}. Nayakshin (2005) showed that an initially
flat accretion disc can be quickly warped due to precession in a
non-axisymmetric gravitational potential. The potential considered there was
that of a stellar ring inclined with respect to the gaseous disc, a situation
which probably existed in the Galactic Centre \citep{NC05,Paumard06}. Finally,
there is in general no reason for an initial disc configuration to be flat
\citep{Phinney89}.

\section{Summary and Conclusions}

In this article I considered some manifestations of stellar activity in the
inner parsecs of galaxies, inside of what is usually considered to be an AGN
accretion disc domain. Stars are important sinks of mass during star formation
and are important radiation and matter sources later on when stars start to
burn their nuclear fuel. Observations of our Galactic Center highlight the
problem faced by accretion disc models on $\ge 0.1$ pc scales. Accretion discs
are very cold there, and one expects these discs to be making stars rather
than accreting. This appears to be exactly what happened with two ring/disks
of gas in the inner $\sim 0.5$ pc of our galaxy about 6 million years ago. If
not for the star formation, Sgr~A$^*$ could be a respectable low-luminosity
AGN now \citep{NC05}. It seems indisputable that a working solution of the
AGN/quasar feeding problem will have to include star formation as well. Much
theoretical work remains to be done in this area.

I also considered the possible role of outflows driven away by stars
populating star forming discs like those observed in Sgr~A$^*$. It is found
that these winds are clumpy, cold and may well contain dust, as required by
the ``torus'' outflow model of Elitzur \& Shlosman (2006). This model thus
might work well for Compton-thin sources. The outflows become optically thick
when they are powered by mass loss rates a factor of $\sim 10$ or more larger
than the mass accretion rate corresponding to Eddington Luminosity for the
AGN. This appears to be a general problem for obscuration by any kind of an
outflow. I propose that warped accretion disks can provide the needed
optically thick AGN obscuration.

\acknowledgements Organisers of the conference are warmly thanked for inviting
me to this splendid and productive meeting, and the opportunity to learn more
about science in China, and China in general.



\end{document}